\documentclass[aps,prb,showpacs,twocolumn,amsmath,amssymb,superscriptaddress]{revtex4}%
\usepackage{graphicx}
\usepackage{dcolumn}
\usepackage{bm}
\usepackage{amsmath}
\usepackage{amsfonts}
\usepackage{amssymb}%

\newcommand{\pa}{\partial}

\newcommand{\be}{\begin{equation}}
\newcommand{\e}{\end{equation}}
\newcommand{\bc}{\begin{cases}}
\newcommand{\ec}{\end{cases}}
\newcommand{\beml}{\begin{subequations}}
\newcommand{\eml}{\end{subequations}}
\newcommand{\beq}{\begin{eqnarray}}
\newcommand{\eq}{\end{eqnarray}}
\newcommand{\ba}{\begin{array}}
\newcommand{\ea}{\end{array}}
\newcommand{\bpm}{\begin{pmatrix}}
\newcommand{\epm}{\end{pmatrix}}
\newcommand{\lt}{\left}
\newcommand{\rt}{\right}
\newcommand{\n}{\nonumber}
\newcommand{\la}{\langle}
\newcommand{\ra}{\rangle}
\newcommand{\ep}{\varepsilon}

\newcommand{\bb}{\boldsymbol}
\newcommand{\h}{^\dagger}

\DeclareMathOperator{\tr}{Tr}

\begin{document}
\title{Spin-Orbit Torques in Two-Dimensional Rashba Ferromagnets}
\author{A.~Qaiumzadeh}
\affiliation{Radboud University, Institute for Molecules and Materials, 6525 AJ Nijmegen, The Netherlands}
\author{R.\,A.~Duine}
\affiliation{Institute for Theoretical Physics and Centre for Extreme Matter and Emergent Phenomena, Utrecht University, 3584 CE Utrecht, The Netherlands}
\author{M.~Titov}
\affiliation{Radboud University, Institute for Molecules and Materials, 6525 AJ Nijmegen, The Netherlands}
\date{\today}
\begin{abstract}
Magnetization dynamics in single-domain ferromagnets can be triggered by a charge current if the spin-orbit coupling is sufficiently strong. We apply functional Keldysh theory to investigate spin-orbit torques in metallic two-dimensional Rashba ferromagnets in the presence of spin-dependent disorders. A reactive, anti-damping-like spin-orbit torque as well as a dissipative, field-like torque is calculated microscopically, to leading order in the spin-orbit interaction strength. By calculating the first vertex correction we show that the intrinsic anti-damping-like torque vanishes unless the scattering rates are spin-dependent.
\end{abstract}
\date{\today}
\maketitle
\section{Introduction}
Spin-orbitronics\cite{spin-orbitronics1,spin-orbitronics2} has attracted a lot of attention recently as a new subfield of spintronics\cite{Spintronics review1,Spintronics review2} in which the relativistic spin-orbit interaction (SOI) plays a central role. Spin-orbitronics includes generation and detection of spin-polarized currents through the spin Hall effect,\cite{SHE-review1,SHE-review2} the induction of non-equilibrium spin accumulations in non-magnetic materials through the Edelstein effect,\cite{edelestein effect1,edelestein effect2} the triggering of magnetization dynamics in single magnetic systems through spin-orbit torques (SOTs),\cite{SOT review1,SOT review2,SOT review3} and magnonic charge pumping by means of inverse SOTs.\cite{Inverse SOT} Spin-orbitronics is believed to ultimately enable the faster and more efficient ways of magnetization switching needed for high density data storage and information processing, thereby providing novel solutions to address the essential challenges of spintronics.
In this paper we investigate the microscopic origin of SOTs in a two-dimensional (2D) metallic ferromagnet with spin-orbit coupling.

The magnetization dynamics in ferromagnets is governed by the seminal Landau-Lifshitz-Gilbert (LLG) equation,\cite{STT review1,STT review2,STT review3}
\be
\label{LLG}
\frac{\pa \bb{m}}{\pa t}=-\gamma\, {\bb{m}} \times {\bb{H}}_{\textrm{eff}}+\alpha_G\, {\bb{m}}\times \frac{\pa {\bb{m}}}{\pa t}+{\bb{T}} ,
\e
where $\bb{m}$ is a unit vector along the magnetization direction $|{\bb{m}}|=1$, $\gamma$ is the gyromagnetic ratio, $\alpha_G$ is the Gilbert damping constant  and ${\bb{H}}_{\textrm{eff}}$ is an effective field which includes the effects of the external magnetic field, exchange interactions, and dipole and anisotropy fields. The first term on the right-hand side of Eq.~(\ref{LLG}) describes the precession of the magnetization vector $\bb{m}$ around the effective field, while the second term describes the relaxation of magnetization to its equilibrium orientation. Furthermore, $\bb{T}$ is a sum of different magnetization torques not contained in the effectieve field or damping.

The spin-polarized current-induced magnetization dynamics in magnetic materials arises as a result of spin transfer torque (STTs).\cite{STT review1,STT review2,STT review3}  It is well known that STT may induce magnetization dynamics in spin-valve structures, and that the exchange interaction between the spin-polarized current and local spins leads, e.g., to domain-wall motion. In uniformly magnetized single-domain systems the transfer of spin angular momentum from the spin-current density $\bb{j}_s$ to a local magnetization is modelled by two different STT terms: i)~an anti-damping-like (ADL) or Slonczewski in-plane torque $\bb{T}\propto \bb{m}\times \bb{m}\times \bb{j}_s$, and ii) an out-of-plane field-like (FL) torque $\bb{T}\propto \bb{m}\times \bb{j}_s$, which is typically negligible in conventional metallic spin valves. On the other hand, in ferromagnets with magnetic domains, in which spin textures such as domain walls are necessarily present, the STT also includes reactive, $\bb{T}\propto\lt(\bb{j}_s\cdot \bb{\nabla}\rt) \bb{m}$, and dissipative, $\bb{T}\propto \bb{m}\times (\bb{j}_s\cdot\bb{\nabla}) \bb{m}$, torques.\cite{STT review1,STT review2,STT review3}

Recently, it was demonstrated both theoretically and experimentally that the current-induced nonequilibrium spin polarization\cite{edelestein effect1,edelestein effect2} in (anti-)ferromagnets with inversion asymmetry may exert a so-called SOT on localized spins and, consequently, may lead to a non-trivial magnetization dynamics.\cite{Garello-switching, SOT1,SOT2,SOT3,SOT4,anti-damping SOT1,anti-damping SOT2,anti-damping SOT-QKE,SOT exp,SOT DW Miron1,SOT DW Miron2,SOT exp Miron1,SOT exp Miron2,SOT exp Miron3,SOT AFM1,SOT AFM2,Linder} Unlike STT, the SOT phenomenon does not require an injection of spin current or the presence of spatial inhomogeneities in the magnetization. The magnetization switching due to SOTs may be achieved with current pulses as short as $\sim 180$\,ps, while the critical charge current density can be as low as $\sim 10^7$\,A\,cm$^{-2}$.\cite{Garello-switching}

Quite generally Rashba SOTs can be classified as either ADL or FL torques \cite{intristic and DW SOT}. The first theoretical and experimental studies of SOT have demonstrated that the ADL SOT is proportional to the disorder strength and can always be regarded as a small correction to the FL SOT.\cite{SOT1,SOT2,SOT3,SOT4, anti-damping SOT1,anti-damping SOT2, anti-damping SOT-QKE, SOT exp} On the other hand, in some recent experiments the opposite statement is made: the torques with ADL symmetry are more likely to be the main source of the observed magnetization behavior.\cite{SOT exp Miron1,SOT exp Miron2,SOT exp Miron3,SHE torque exp,SHE-SOT-thickness exp1,SHE-SOT-thickness exp2,SOT exp Yaroslav} These experiments are performed with ferromagnetic metals grown on top of a heavy metal with strong SOI and may, in principle, be explained by the spin Hall effect which induces a spin-polarized current. This spin current, in turn, exerts a torque on the magnetic layer via the STT mechanism\cite{SHE-SOT-theo, SHE-SOT-thickness exp1,SHE-SOT-thickness exp2} so that the ADL symmetry term plays the major role in the effect as discussed above.

It is, however, a serious experimental challenge to distinguish between SOT and spin-Hall STT in bilayers, since both torques have the same symmetry.\cite{SHE-SOT-thickness exp1,SHE-SOT-thickness exp2} Very recently Kurebayashi \textit{et al.}\cite{intristic SOT exp1} conducted an experiment on the bulk of strained GaMnAs, which has an intrinsic crystalline asymmetry. In these experiments, the contribution of a possible spin-Hall-effect STT was completely eliminated, while sizable ADL torques were nevertheless detected. This provides a strong argument in favour of the ADL-SOT nature of the observed torque. The authors of Ref.~\onlinecite{intristic SOT exp1} attribute this torque to an intrinsic Berry curvature, and estimate a scattering-independent, i.e. intrinsic, ADL-SOT.\cite{intristic SOT exp1,intristic SOT exp2,Blugel} This intrinsic ADL-SOT has also been reported by van der Bijl and Duine.\cite{intristic and DW SOT}

In this paper we calculate both FL- and ADL-SOTs in a 2D Rashba ferromagnetic metal microscopically by using a functional Keldysh theory approach.\cite{functional keldysh} By calculating the first vertex correction we show that the intrinsic ADL-SOT vanishes unless the impurity scattering is spin dependent.

The rest of this paper is organized as follows. Section II introduces the model and method. In Sec. III we calculate
SOTs with and without vertex corrections. We conclude our work in Sec. IV.
\section{Model Hamiltonian and method}
We start with the 2D mean-field Hamiltonian ($\hbar=c=1$),
\be
\label{Ham}
\mathcal{H}[\psi\h, \psi]=\int d^2\bb{r}\;\psi\h_{\bb{r},t} \lt[H_0+V_{\textrm{imp}}+\hat{\bb{j}}\cdot\bb{A}_t\rt] \psi_{\bb{r},t}.
\e
where ${\bb{\psi}}^\dag=({\bb{\psi}}_\uparrow^*,{\bb{\psi}}_\downarrow^*)$ is the Grassman coherent state spinor. Here, $H_0$ is the 2D conducting ferromagnet Hamiltonian density in the presence of Rashba SOI,\cite{Rashba review}
\be
\label{H0}
H_0= \frac{{\bb{p}}^2}{2m_e}+\alpha_R\,(\bb{\sigma}\times\hat{\bb{z}})\cdot\bb{p} -\tfrac{1}{2}\Delta\,\bb{\sigma}\cdot\bb{n}_{\bb{r},t}-\tfrac{1}{2}\Delta_\textrm{B} \sigma_z,
\e
where $\bb{p}$ is the 2D momentum operator, $\alpha_R$ is the strength of the SOI, $\Delta$ and $\Delta_{B}$ are the exchange energy and the Zeeman
splitting due to an external field in the $z$-direction, respectively, ${\bb{n}}_{{\bf{r}},t}$ is an arbitrary unit vector that determines the quantization axis, and $\bb{\sigma}$ is the three-dimensional vector of Pauli matrices.

The vector potential $\bb{A}_t= \bb{E}e^{-i\Omega t}/i\Omega$ is included in Eq.~(\ref{Ham}) to model a $dc$ electric field in the limit $\Omega\to 0$. It is coupled to the current density operator, which is given by $\hat{\bb{j}}=(ie/2m_e)(\overleftarrow{\bb{\nabla}} -\overrightarrow{\bb{\nabla}}) -e \alpha_R\,\bb{\sigma}\times\hat{\bb{z}}$, where $e$ is the electron charge and $m_e$ is the electron effective mass.  Finally, the impurity potential $V_{\mathrm{imp}}$ is of the form
\be
V_{\mathrm{imp}}(\bb{r})=\bpm V_\uparrow & 0 \\ 0 & V_\downarrow \epm
\sum_{i}\delta(\textbf{r}-\textbf{R}_i),
\e
where $V_{\uparrow(\downarrow)}$ is the strength of spin-up (down) disorder, and the index $i$ labels the impurity centers $\textbf{R}_i$. More specifically, we restrict ourselves to the gaussian limit of the disorder potential.

The impurity-averaged retarded Green's function in the Born approximation is given by\cite{green function1,green function2,green function3}
\be
G^{+}_{{\bb{k}},\ep}=\lt(g^{-1}_\downarrow \sigma^\uparrow+g^{-1}_\uparrow\sigma^\downarrow+\alpha_R(\sigma_yk_x-\sigma_xk_y)\rt)^{-1}.
\e
where $g^{-1}_s=\ep -\ep_k+ s M+i\gamma_s$, for $s=\uparrow\!\!\!(+)$ or $\downarrow\!\!\!(-)$, $\sigma^s=(\sigma_0+s \sigma_z)/2$, ${\bb{k}}$ and $\ep$ are the wavevector and energy, respectively, $\ep_k=k^2/2m_e$, and $M=(\Delta+\Delta_{B})/2$. We have also introduced the spin-dependent scattering rate $\gamma_{s}=\pi \nu n_{\mathrm{imp}}V^2_{s}$, where $\nu_0=m_e/2\pi$ is the density of states per spin for 2D electron gas, and $n_{\mathrm{imp}}$ denotes the impurity concentration. Here we have assumed that both spin-orbit split bands are occupied, i.e. the Fermi energy is larger than magnetization splitting, $\ep_F> M$.

Following Ref.~\onlinecite{functional keldysh} we minimize the effective action on the Keldysh contour with respect to quantum fluctuations of $\bb{n}$. This procedure gives us directly the LLG equation which contains torque terms in linear response with respect to the external field $\bb{E}$. The effective action is given by $S=\int_{\mathcal{C}^K} dt L_F(t)$, where $\mathcal{C}^K$ stands for the Keldysh contour and $L_F(t)=\int d^2\bb{r}(\hat{\bb{\psi}}\h_{\bb{r},t}i \frac{\pa}{\pa t}\hat{\bb{\psi}}_{\bb{r},t}-\mathcal{H})$ is the mean-field Lagrangian.

We further assume that we are dealing with a ferromagnetic metal which is uniformly magnetized in the $z$-direction. Thus, we can approximate the vector $\bb{n}$ as
\be
\bb{n}_{\bb{r},t}\simeq
\bpm \delta n^x_{{\bb{r}},t} \\ \delta n^y_{{\bb{r}},t} \\ 1-\frac{1}{2}(\delta n^x_{{\bb{r}},t})^2-\frac{1}{2}(\delta n^y_{{\bb{r}},t})^2) \\
\epm.
\e
In order to derive the LLG equation with torque terms it is sufficient to expand the effective action up to second order in $\delta \bb{n}$ and up to  first order in the vector potential: $S_{\mathrm{eff}}=S_{\mathrm{SOT}}[\mathcal{O}(\delta {\bf{n}}), \mathbf{A}]+ S_{\mathrm{rest}}[\mathcal{O}(\delta {\bf{n}}^2),\mathbf{A}=0]$. A straightforward calculation gives
\be
S_{\mathrm{SOT}}
=\int_{\mathcal{C}^K}\!\!\!dt\int_{\mathcal{C}^K}\!\!\!dt'\int \!d^2{\bb{r}}
\int \!d^2{\bb{r}}'\, {\chi}_{a;\bb{r}-\bb{r}';t,t'}
\delta n^a_{{\bb{r}'},t'},
\e
where $\chi_{a}$ ($a=\{x,y\}$) is the response function,
\be
{\chi}_{a;\bb{r}-\bb{r}';t,t'}=\frac{i\Delta}{4}\lt\la {\bb{j}}_{\bb{r},t}\;\psi\h_{\bb{r}',t'}\sigma_a\psi_{\bb{r}',t'}\rt\ra \cdot \bb{A}_{t}, \label{response-function}
\e
and $\bb{j}=\psi\h\hat{\bb{j}}\psi$ is the charge current density.
Note that in the absence of SOI the term $S_{\mathrm{SOT}}$ is 0 and only second order terms, $S_{\mathrm{rest}}$ \cite{functional keldysh}, remain.
The field $\delta \bb{n}$ can be split into the physical magnetization field $\delta {\bf m}$ and a quantum fluctuation field $\bm{\xi}$ as $\delta n^a_{\bb{r},t_{\pm}}=\delta m^a_{\bb{r},t}\pm\xi^a_{\bb{r},t}/2$, where $+$ corresponds to the upper and $-$ to the lower branch of the Keldysh contour.  At first order with respect to the quantum component we obtain
\be
S_{\mathrm{SOT}}=\int\! dt\int\! dt'\!\int\! d^2{\bb{r}}'\!\! \int\! d^2{\bb{r}}\; \chi^{-}_{a;\mathbf{r}-\mathbf{r}';t,t'}\xi^a_{\mathbf{r}',t'},
\e
where $\chi^{-}$ is the advanced component of the correlator, and the sum over repeated indices $a$ is assumed.
The LLG equation is, then, derived by minimizing the effective action with respect to quantum fluctuations, $\delta S_{\mathrm{eff}}/\delta \bb{\xi}=0$. Thus, the transverse components of the LLG equation in the Fourier space are given by,
\be
\label{LLG0}
\mathcal{F}\lt[\frac{\delta S_{\mathrm{rest}}}{\delta\xi_a}\rt]_{{\bf{q}}=0,\ep}+\chi^{-}_{a;{\bf{q}}=0,\ep=0}=0,
\e
where $\mathcal{F}[...]$ represents the Fourier transformation operator. The functional derivative in Eq.~(\ref{LLG0}) gives the precession and Gilbert
damping terms of the LLG equation,\cite{functional keldysh} while the second term describes the SOT. The dependence of Gilbert damping on SOI is second order in $\alpha_R$,\cite{intristic and DW SOT} and we focus below on SOT which is of first order in $\alpha_R$. The appearance of the zero-momentum response function $\chi_{a;{\bf{q}}=0,\ep=0}$ in the LLG equation shows that the SOT is finite even for spatially uniform magnetization, in contrast to the (non-)adiabatic STT which is of the first order in the gradient of magnetization.

\begin{figure}
\centering
\includegraphics[width=6cm]{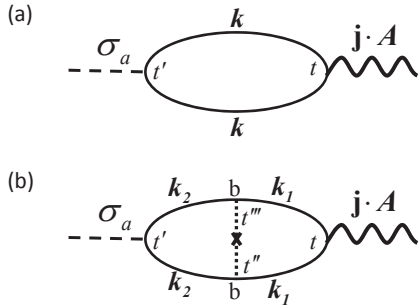}
\caption{Feynman diagrams related to the spin-torque response function Eq. (\ref{response-function}): (a) undressed response function, and (b) the first vertex correction. The solid line corresponds to an electron propagator in the Born approximation, the wiggly line to the coupling to vector potential and current, and the dashed represents a spin fluctuation. The vertical dotted line describes the averaging over impurity positions.}\label{diagram}
\end{figure}

\section{Calculation of SOTs}

In what follows we evaluate the spin-torque response function of Eq.~(\ref{response-function}), shown diagrammatically in Fig.~\ref{diagram}, to derive the SOT in the ballistic limit $\gamma_s \ll k_B T$, where $k_B T$ is the thermal energy. We calculate first the bare (undressed) part of the response function, $\chi^{(0)}$, depicted in Fig.~\ref{diagram}a. The final result for spin torque is, then, obtained by adding the first vertex correction, $\chi^{(1)}$, depicted in Fig.~\ref{diagram}b. Throughout the calculation we assume that $\gamma_s \ll k_B T \ll \alpha_R k_F\ll M$, where $k_F$ is the Fermi wavevector. The condition $ \alpha_R k_F\ll M$ is normally fulfilled in the metallic ferromagnets of interest. Whether or not the condition $\gamma_s\ll k_B T\ll \alpha_Rk_F$ is fulfilled depends strongly on the sample quality.  The analysis of spin torques in diffusive regime $k_B T\ll \gamma_s$ will require calculation of the full vertex correction and will be done elsewhere.

\subsection{Undressed response function}
The spin-torque response function of Eq.~(\ref{response-function}) without vertex corrections is given by
\be
\label{chi0}
{\chi}^{(0)}_{a;t,t'}=\frac{e\Delta}{4i} \int\frac{d^2\bb{k}}{(2\pi)^2}\,\mathrm{Tr}[\bb{v}_{\bb{k}}\check{G}_{{\bb{k}};t,t'}\sigma_a\check{G}_{{\bb{k}};t',t}]\cdot\mathbf{A}_{t}.
\e
where $\bb{v}_{\bb{k}}=\bb{k}/m_e-\alpha_R \bb{\sigma}\times \hat{\bb{z}}$ is the velocity vector, and $\check{G}$ is the Green's function on the Keldysh contour. From Eq.~(\ref{chi0}) we find retarded and advanced components of the response function in the limit of zero frequency and momentum as
\beq
\label{chi0pm}
{\chi}^{(0)\pm}_{a}
&=&\frac{e\Delta}{4i} \lim_{\Omega\rightarrow 0}\int\!\frac{d^2{\bb{k}}}{(2\pi)^2}
\int\!\!\!\!\int d\omega\,d\omega'\, \frac{f_{\omega'}-f_{\omega}}{\Omega+\omega'-\omega\pm i0} \n \\
&\times&
\frac{1}{\Omega}\mathrm{Tr}[({\bb{v}}_{{\bb{k}}}\cdot\bb{E})\mathcal{A}_{{\bb{k}},\omega}\sigma_a
\mathcal{A}_{{\bb{k}},\omega'}],
\eq
where $\mathcal{A}_{{\bb{k}},\omega}=i(G^{+}_{\bb{k},\omega}-G^{-}_{\bb{k},\omega})/2\pi$ is the spectral function and $f_\omega=[e^{(\omega - \epsilon_F)/k_B T}+1]^{-1}$ stands for the Fermi distribution function.

In the limit of weak disorder, we can decompose the response function into two parts: the intrinsic part $\chi_{\textrm{in}}$, which turns out not to depend on the scattering rate and describes interband transitions and the extrinsic part $\chi_{\textrm{ex}}$, which essentially depends on disorder and corresponds to intraband contributions. The intrinsic part corresponds to the principal value integration in Eq. (\ref{chi0pm}), while the extrinsic part is given by the corresponding delta-function contribution. To leading order in $\alpha_R$ we find
\beml
\beq
\label{inrinsic-chi}
{\chi}^{(0)-}_{\textrm{in},a} &=& \frac{e\alpha_R\Delta}{8 M}  \nu_0 E_{a},\\
{\chi}^{(0)-}_{\textrm{ex},a} &=&\frac{e\alpha_R\Delta}{8 M}\nu_0\lt[\frac{\ep_F\!-\!M}{\gamma_\downarrow}\!-\!\frac{\ep_F\!+\!M}{\gamma_\uparrow}\rt](\hat{\bb{z}}\times {\bb{E}})_a.\qquad
\label{extrinsic-chi}
\eq
\eml
The corresponding expressions for the SOTs are the ADL $\mathbf{T}^{\mathrm{ADL}}$ and FL $\mathbf{T}^{\mathrm{FL}}$ contributions, which do not take into account vertex corrections,
\beml
\beq
\bb{T}^{(0)}_{\mathrm{ADL}}&=& -2e\alpha_R \nu_0 \bb{m}\times\bb{m}\times(\hat{\bb{z}}\times\bb{E}),
\label{intrinsic-SOT}\\
\bb{T}^{(0)}_{\mathrm{FL}}&=& -\frac{e\alpha_R  \Delta \nu_0}{M}
\lt[\!\frac{\ep_F\!+\!M}{\gamma_\uparrow}-\frac{\ep_F\!-\!M}{\gamma_\downarrow}\!\rt]\bb{m}\times(\hat{\bb{z}}\times\bb{E}).\qquad
\label{extrinsic-SOT}
\eq
\eml
Hence, we find that the ADL SOT in the absence of vertex corrections has an intrinsic origin, i.e., is disorder-independent.

\subsection{Vertex correction}
Let us now turn to the calculation of the first vertex correction to the spin-torque response function depicted in Fig.~\ref{diagram}b. For the corresponding response function on the Keldysh contour we find
\beq
&&\chi^{(1)}_{a;t,t'}=
\frac{e\Delta}{4i}\int\frac{d{\bb{k}}_1}{(2\pi)^2} \int\frac{d{\bb{k}}_2}{(2\pi)^2}\int_{c^K}\!\!\! dt_1\int_{c^K}\!\!\! dt_2
\tr[\mathbf{A}_{t}\cdot{\bb{v}}_{{\bb{k}}_1} \n \\
&&\times \check{G}_{\bb{k}_1;t,t_1}\langle V_{\mathrm{imp}}
\check{G}_{\bb{k}_2;t_1,t'}\sigma_a\check{G}_{{\bb{k}}_2;t',t_2}V_{\mathrm{imp}}\rangle \check{G}_{{\bf{k}}_1;t_2,t}].
\eq
The advanced component of $\chi^{(1)}$ at zero energy and momentum is, then, given by
\beq
\chi^{(1)-}_{a}\!\!\!&=&\frac{e\Delta}{4i} \eta_b\!\int\!\!\frac{d^2\bb{k}_1}{(2\pi)^2}\!\int\!\!\frac{d^2\bb{k}_2}{(2\pi)^2}\!
\int\!\!\!\!\int \!\!d\omega\,d\omega'\, \frac{f_{\omega'}-f_{\omega}}{\Omega\!+\!\omega\!-\!\omega'\!-\!i0} \n \\
&\times&\frac{1}{\Omega}\mathrm{Tr}\big[\bb{E}\cdot{\bb{v}}_{{\bf{k}}_1}\big(G^{+}_{{\bb{k}}_1,\omega}\sigma_{b}\mathcal{A}_{{\bb{k}}_2,\omega}\sigma_aG^{+}_{{\bb{k}}_2,\omega'}\sigma_{b}\mathcal{A}_{{\bb{k}}_1,\omega'}\n\\
&&\qquad+\,G^{+}_{{\bb{k}}_1,\omega}\sigma_{b}\mathcal{A}_{{\bb{k}}_2,\omega}\sigma_a\mathcal{A}_{{\bb{k}}_2,\omega'}\sigma_{b}G^{-}_{{\bb{k}}_1,\omega'}\n\\
&&\qquad+\,\mathcal{A}_{{\bb{k}}_1,\omega}\sigma_{b}G^{-}_{{\bb{k}}_2,\omega}\sigma_aG^{+}_{{\bb{k}}_2,\omega'}\sigma_{b}\mathcal{A}_{{\bb{k}}_1,\omega'} \n\\
&&\qquad+\,\mathcal{A}_{{\bb{k}}_1,\omega}\sigma_{b}G^{-}_{{\bb{k}}_2,\omega}\sigma_a\mathcal{A}_{{\bb{k}}_2,\omega'}\sigma_{b}G^{-}_{{\bb{k}}_1,\omega'}\big)\big],
\eq
where the summation over the index $b=\{0,z\}$ and the limit $\Omega \to 0$ are assumed. We have also used the notations $\eta_0=n_{\mathrm{imp}}(V_\uparrow+V_\downarrow)^2/4$ and $\eta_z=n_{\mathrm{imp}}(V_\uparrow-V_\downarrow)^2/4$. Using the same approximations as for the undressed part of the response function we obtain the intrinsic contribution as
\be
\chi^{(1)-}_{\textrm{in},a} = -\frac{e\alpha_R\Delta}{8 M}\nu_0\frac{\gamma_\uparrow+\gamma_\downarrow}{2(\gamma_\uparrow
\gamma_\downarrow)^{\frac{1}{2}}}E_{a},
\e
while the corresponding extrinsic contribution is of the second order in scattering rates and can be neglected.

Thus, we obtain the FL and ADL torques in the limit $\gamma_s\ll\alpha_Rk_F\ll M$ to the leading order in the SOI as
\beq
\bb{T}_{\mathrm{FL}}\!\! &= &\!\!\frac{e\alpha_R \nu_0 \Delta}{M}\lt[\frac{\ep_F-M}{\gamma_\downarrow}-\frac{\ep_F+M}{\gamma_\uparrow}\rt]
\bb{m}\!\times\!(\hat{\bb{z}}\!\times\!\bb{E}),\label{FL-SOT}\\
\bb{T}_{\mathrm{ADL}}\!\!&= &\!\!\lt[\frac{\gamma_\uparrow+\gamma_\downarrow}{2\sqrt{\gamma_\uparrow
\gamma_\downarrow}}-1\rt]2e\alpha_R \nu_0\bb{m}\!\times\!\lt(\bb{m}\!\times\!(\hat{\bb{z}}\!\times\!\bb{E})\rt).\qquad
\label{ADL-SOT}
\eq
These expressions provide the main result of this paper.

\section{Conclusions}
The SOT mechanism is based on the exchange of angular momentum between the crystal lattice and the local magnetization via spin-orbit coupling. Here, we found the FL- and ADL-SOTs microscopically, Eqs. (\ref{FL-SOT}) and (\ref{ADL-SOT}). The FL-SOT originates from the Fermi surface contribution of the response function Eq. (\ref{response-function}), while the ADL-SOT is acquires contributions from the entire bands. Our main result in Eq. (\ref{ADL-SOT}) immediately shows that the intrinsic contribution to ADL-SOT is completely canceled in the presence of spin-independent scattering $\gamma_\uparrow=\gamma_\downarrow$. That is, the intrinsic
component of the ADL SOT, which originates from virtual interbranch
transitions, is canceled by the vertex correction when weak spin-independent impurity scattering is taken into account. Our result, therefore, explicitly elucidates the interplay between intrinsic and extrinsic contributions to ADL SOT. This result resembles the suppression of both spin Hall conductivity in nonmagnetic metals and anomalous Hall conductivity in magnetic metals, in the presence of spin-independent disorder.\cite{green function1,green function2,green function3,Gerrit-AHE,Sakai}  In these effects the cancelation is model dependent, and occurs for parabolic band dispersion and linear-in-momentum SOI. We expect a similar scenario for intrinsic SOT.

The existence of a Rashba effect on the interface between an ultrathin ferromagnet and a heavy metal is the subject of intense discussion. Our results show that the amplitudes of the FL and ADL SOTs can be of the same order of magnitude depending on the relative strengths of the SOI, spin-dependent scattering rates, and exchange interaction. Our results may qualitatively describe Co/Pt interfaces which are characterised by particularly large Rashba SOI of the magnitude of $1$\,eV\,{\AA}.\cite{SOT exp Miron1,SOT exp Miron2,SOT exp Miron3, rashba strength}  Relating the strength of the Rashba coupling to the magnitude of the SOTs, however, would require ab-initio modeling and additional experimental information. Which of our results apply to more general models and band structures will be the subject of future investigation.

\section*{ACKNOWLEDGEMENT}
We acknowledge Hiroshi Kohno and Dmitry Yudin for useful discussions. The work was supported by Dutch Science Foundation NWO/FOM 13PR3118 and by EU Network FP7-PEOPLE-2013-IRSES Grant No 612624 "InterNoM". R.D. is a member of the D-ITP consortium,
a program of the Netherlands Organisation for Scientific Research (NWO), which is funded by the Dutch Ministry of Education, Culture and Science (OCW).

\end{document}